Magnetism in Cobalt doped $Cu_2O$ thin films without and with Al, V, Zn codopants.

S. N. Kale[1,*], S. B. Ogale[1], S. R. Shinde[1], M. Sahasrabuddhe[2], V. N. Kulkarni[1], R. L. Greene[1], and T. Venkatesan[1]

[1]Center for Superconductivity Research, Department of Physics, University of Maryland, College Park, MD20740-4111.

[2]Department of Physics, University of Poona, Pune 411 007, India.

Thin films of 5 % Co doped $Cu_2O$ were grown on single crystal (001) MgO substrates by pulsed laser deposition, without and with 0.5 % codoping with Al, V or Zn. Structural, electrical, and magnetic properties were studied. The films showed phase pure character under the chosen optimum growth conditions. Spin glass like behavior was observed in Co doped films without codoping. A clear ferromagnetic signal at room temperature was found only in the case of Co:$Cu_2O$ films codoped with Al.

* On leave from Fergusson College, Pune, India.

Diluted Magnetic Semiconductors (DMS) are materials of great interest at the present time because of their projected potential for the rapidly evolving field of spintronics[1-5]. Early efforts in this field focused on compound semiconductor materials doped with transition metal atoms, and ferromagnetism was indeed realized in some cases. The Curie temperatures in these systems have however been rather low (e.g. 110 K for $Ga_{1-x}Mn_xAs$), prompting searches in other classes of materials. In this context, oxides are a natural class to explore in view of their broad range of chemically tunable properties. Search for magnetism in transition element doped ZnO and $TiO_2$ has shown considerable promise[6-11], although the precise origin of ferromagnetism in these systems and the specific microstate of the materials are issues being currently debated in the literature. While the main focus of discussion in this context has been on the carrier-induced ferromagnetism, other mechanisms such as that based on percolation of bound magnetic polarons[12] have also been proposed. The latter type of proposals could even be applicable to insulating states[12,13].

In this paper we explore the possibility of inducing ferromagnetism in Cuprous Oxide ($Cu_2O$) by dilute cobalt doping. In addition to doping with 5 % Co, we have also examined cases of codoping with 0.5 % Al, V, or Zn, (which bear different valence states) in an attempt to influence the magnetism through possible carrier concentration and defect state changes. Cuprous oxide can be grown epitaxially on (001) MgO by pulsed laser deposition[14]. This material is an insulator with a bandgap of about 2 eV and room temperature resistivity in the range $10^2$ to $10^6$ $\Omega$-cm[15-17]. This value is seen to be highly influenced by the deposition technique. Cuprous oxide is useful as an energy converter for solar cell applications[18], and as humidity and gas sensor material[19,20]. It is an attractive material because it has advantages of non-toxicity, high absorption coefficient, and low production cost[21].

The ceramic targets of $Co_{0.05}R_{0.005}Cu_{0.945}O$ (where R= Al, V, Zn) used for pulsed laser ablation were synthesized by the standard solid-state reaction technique. Targets of pure CuO and $Co_{0.05}Cu_{0.95}O$ were also similarly prepared for a comparative study. Based on the guidance of the Cu-O phase diagram and the previous study on $Cu_2O$ film growth[14], the depositions were performed at a substrate temperature of 700 °C and oxygen partial pressure of 1 x $10^{-3}$ Torr. The laser energy density and pulse repetition rate were kept at 1.8 J/cm$^2$ and 10 Hz, respectively. The samples were cooled in the same pressure as that used during deposition, at the rate of 20 °C/min. MgO (001) substrates were used since its lattice parameter (4.213 Å) matches closely with that of $Cu_2O$ (4.296 Å). Prior to deposition, the MgO substrates were etched with hot phosphoric acid to yield good quality epitaxial films[14]. The films were characterized by x-ray diffraction (XRD), SQUID magnetometry, and transport measurements.

Fig. 1(a) shows the XRD pattern (log scale) for the undoped $Cu_2O$ film grown on (001) MgO. These data reveal the good structural quality of the film with a high degree of orientational order. The same was found to be the case for Co-doped and Al, Zn, V codoped films. The corresponding XRD patterns for the primary (200) peak are shown in Fig. 1(b); the patterns being shifted along the y-axis for clarity. It is useful to mention here that high quality $Cu_2O$ films could also be grown on R-plane sapphire substrates, but with a (110) orientation. Using both side polished sapphire the optical properties of $Cu_2O$ film could be evaluated and were also found to be good. In the inset to Fig. 1 is shown a plot of $(\alpha E)^{1/2}$ vs. photon energy(E), which reveals that the optical bandgap of the film is about 2.05 eV, as expected. These structural and optical data together ensure the goodness of the chosen growth conditions.

In Fig. 2 we show the magnetization as a function of temperature for the Co-doped $Cu_2O$ film without and with Al, V and Zn codoping, measured from 4.2 K to 300 K using a SQUID magnetometer. It can be seen that doping $Cu_2O$ only with Co leads to a spin glass like behavior with some peculiar features at 170 and 250 K, the origin of which is not clear at present and would require further studies. Zn and V codoping not only seem to supress this spin glass behavior, but also do not lead to any discernable ferromagnetic signature. Significant magnetization at room temperature is only seen in the case of Co doped $Cu_2O$ that is codoped with 0.5% Al. Even in the Al codoped sample a rise in magnetization is seen below about 50 K. The hysterisis loop obtained at room temperature for the Al codoped Co:$Cu_2O$ sample is shown in the inset. Appearance of a well-defined loop with a coercivity of about 50 Oe signifies ferromagnetism.

Fig. 3 shows the resistivity data as a function of temperature for various samples. This resistivity measurement was performed in a current-in-plane (CIP) geometry and hence may not truly represent the bulk resistivity alone. The growth of $Cu_2O$ on MgO is suggested to occur in the form of coherent islands[22]. Hence a current-perpendicular-to-plane (CPP) measurement may be needed to elucidate the influence of dopants on the intragrain transport. Such measurements need a conducting bottom electrode, which would not affect the growth. This work is currently in progress. The CIP room temperature resistivity of the pure $Cu_2O$ film is seen to be quite high ( 225 Ω-cm) as expected for a fairly high quality material[15-17]. With 5 % Co doping the resistivity value increases to 512 ohm-cm. Codoping of Al with Co does not cause a significant change in resistivity, but vanadium doping increases the resistivity at 300K to about 1800 ohm-cm. Interestingly, the room temperature resistivity goes down to 46 ohm-cm, for the case of Zn codoping.

The fact that the room temperature ferromagnetic signal was not seen for a low resistivity Zn codoped Co:$Cu_2O$ sample, but was seen for a relatively resistive Al codoped Co:$Cu_2O$ sample implies that the mechanism for occurrence of ferromagnetism in this system is most possibly not carrier induced. The possibility of pure Co metal clusters also seems unlikely based on the rather low value (0.44 $\mu_B$/Co) of the observed moment per Co, which is much smaller than that for Co (1.67 $\mu_B$/Co) or for Co nanoclusters (2.1 ± 0.5 $\mu_B$/Co). Kaminski et al.[12] have discussed a mechanism based on percolation of bound magnetic polarons that is not RKKY type, and the same may be applicable in this case. Interestingly, Zn and V are 3d transition elements with orbitals compatible with those of Cu and Co which are also 3d elements. Zn holds a fixed valence state of 2+ while V can support mixed valence. Neither of these codopants however seem to induce a ferromagnetic state in the Co doped $Cu_2O$. On the other hand Al, which not only has the smallest ionic radius amongst the codopants but also s and p as the outermost orbitals and therefore no orbital compatibility with the 3d elements, does induce ferromagnetism. If one views at this as an orbital defect state, it is possible that the corresponding disorder is responsible for ferromagnetism. The idea of defect mediated ferromagnetism has already been applied to Mn doped $CdGeP_2$ system[23], which is claimed to exhibit room temperature ferromagnetism[24]. Further experiments and theoretical insights are clearly needed to sort out these issues.

In conclusion, ferromagnetism at room temperature is observed in 5% Co doped $Cu_2O$ films only with codoping with 0.5% Al. A clear hysterisis loop with coercivity of about 50 Oe is seen. Codoping with Zn or V does not lead to ferromagnetism but causes changes in resistivity. Absence of a clear correlation with resistivity, and appearance of ferromagnetism only under Al codoping suggests that the mechanism may be related to orbital defects.

This work was supported under DARPA (grant # N000140210962) and NSF-MRSEC 00-80008.

Figure Captions

Fig. 1: (a) XRD pattern (log scale) for the undoped $Cu_2O$ film grown on (100) MgO. The inset shows the plot of $(\alpha E)^{1/2}$ vs. photon energy (E) which reveals that the optical bandgap of the film is about 2.05 eV. (b) XRD pattern for Co doped and Al, Zn, V codoped films. The patterns being shifted along the y-axis for clarity.

Fig. 2: (a) Plot of magnetization as a function of temperature for the Co doped $Cu_2O$ film without and with Al, V and Zn codoping, measured from 4.2 K to 300 K using SQUID magnetometer. The inset shows room temperature hysteresis for the Al codoped $Co:Cu_2O$ sample. A well defined loop with a coercivity of about 50 Oe signifies ferromagnetism.

Fig. 3: The resistivity data as a function of temperature for various samples. The room temperature resistivity for undoped film is 225 ohm-cm, which increases to 512 ohm-cm with Co doping. Upon codoping with Al, V and Zn, the resistivity changes to 500, 1800 and 46 ohm-cm respectively.

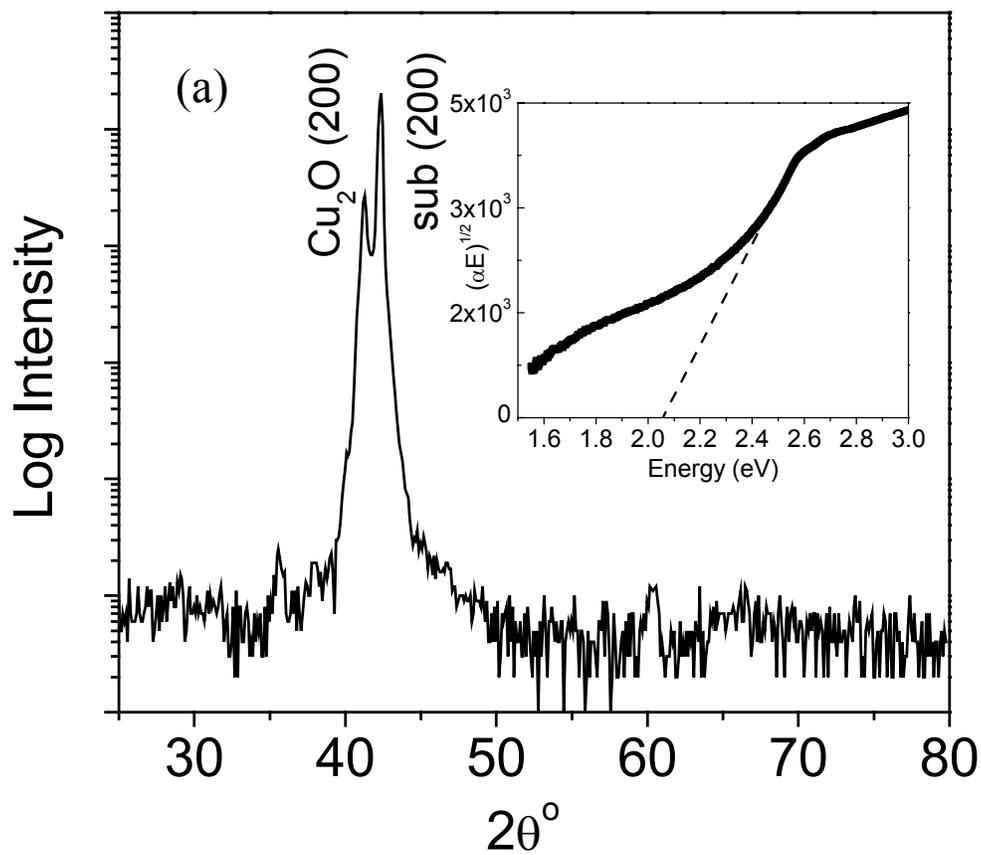

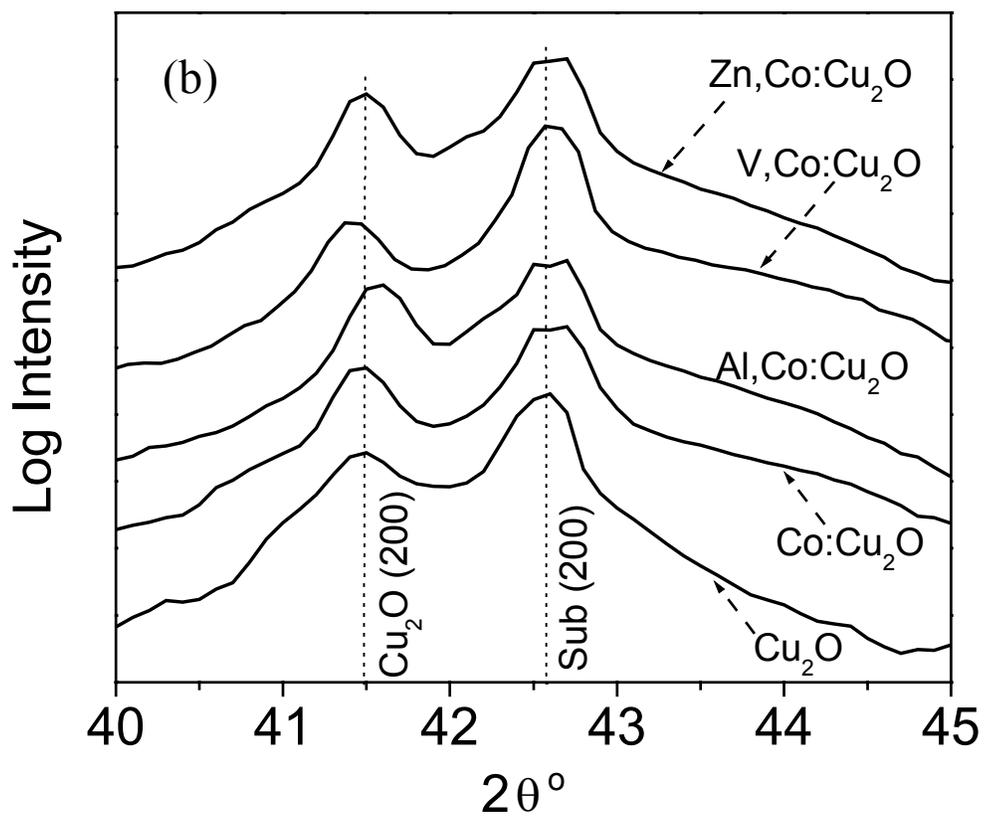

Fig.1 of Kale et al.

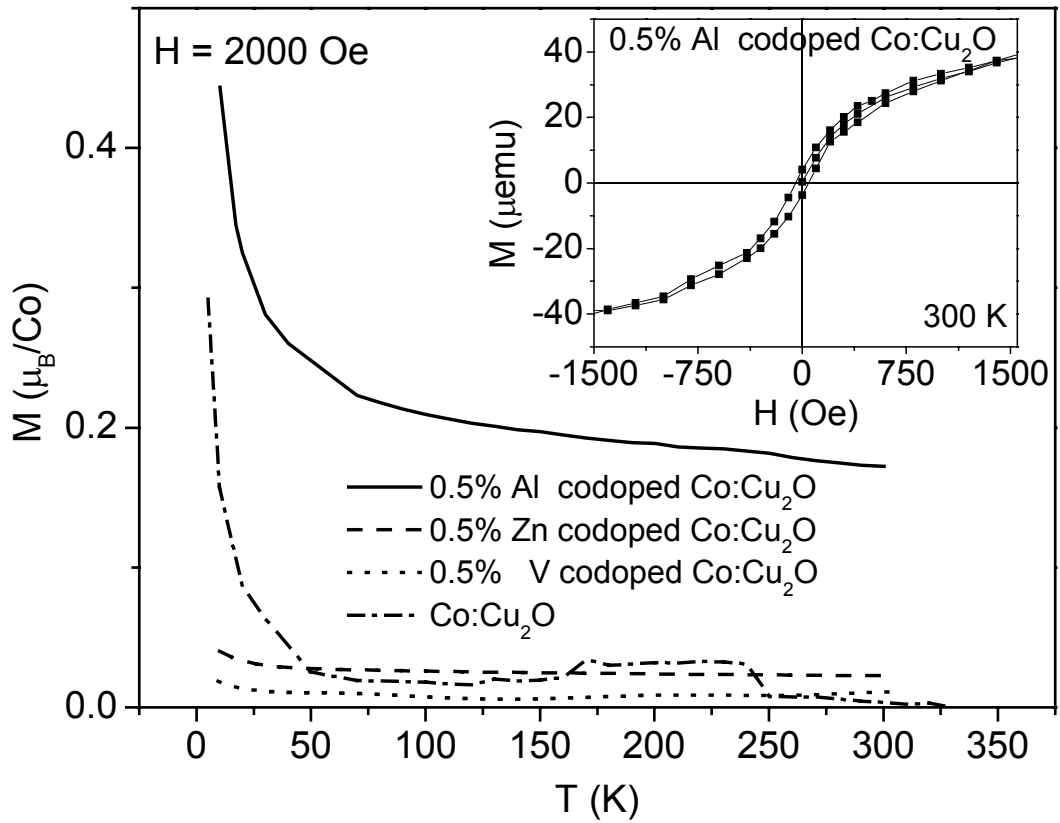

Fig.2 of Kale et al.

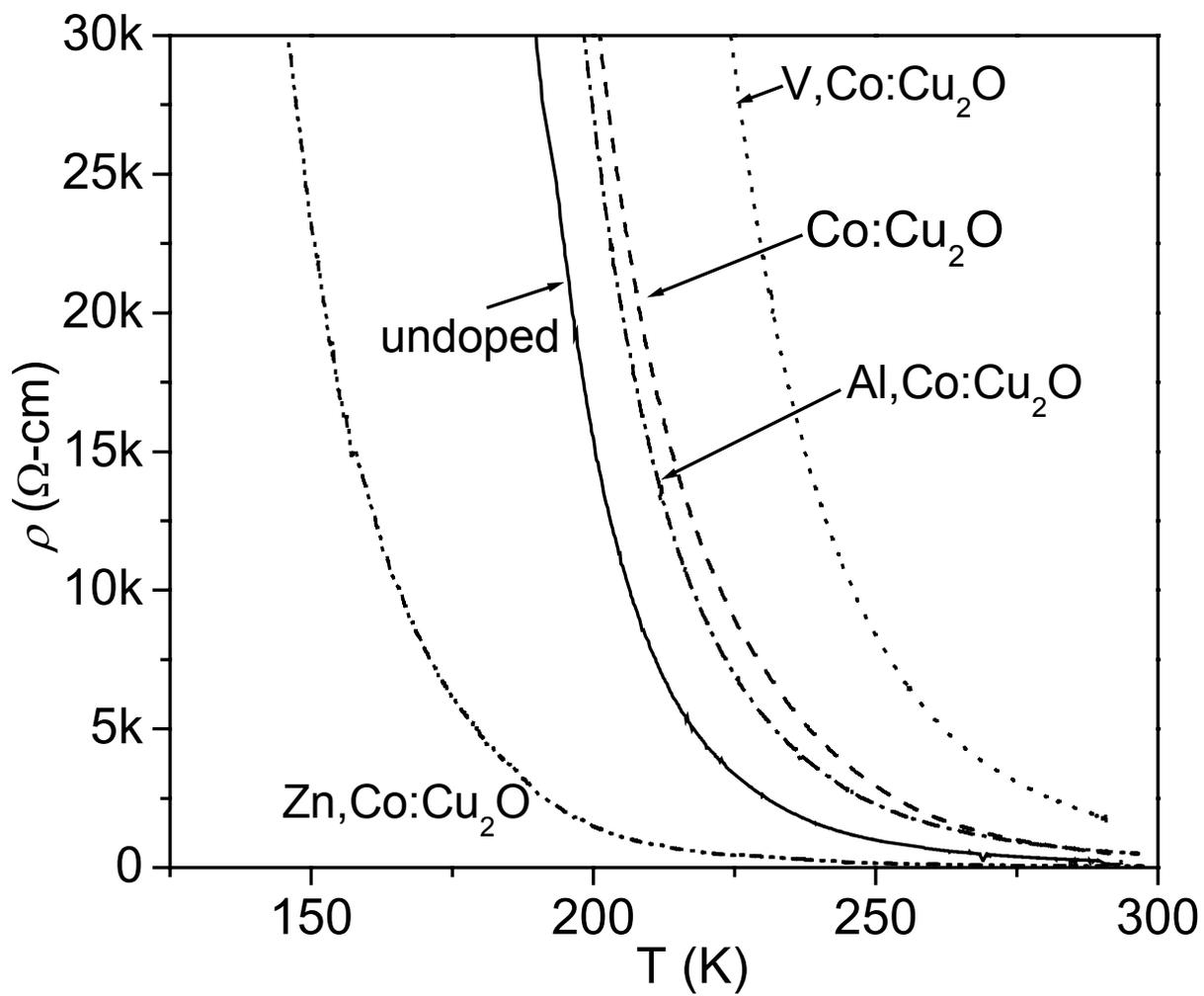

Fig.3 of Kale et al.